\documentclass{jpsj3}
\usepackage{amsmath}
\usepackage{graphicx}
\usepackage{amssymb}

\usepackage{gt21jpsj}
\usepackage{color} 
\usepackage{graphicx}
\usepackage{bm}
\usepackage{amsmath}

\def\newblock{\hskip .11em plus .33em minus .07em} 
\begin{document}
\title{
Spin Hall response at finite wave vector in ferromagnets 
}
\author{Gen Tatara} 
\inst
{RIKEN Center for Emergent Matter Science (CEMS)
and RIKEN Cluster for Pioneering Research (CPR), 
2-1 Hirosawa, Wako, Saitama, 351-0198 Japan}
\date{\today}
\abst{
Spin Hall effect at finite wave vector in a ferromagnetic conductor is theoretically studied by calculating the spin density as the linear response to an applied electric field.
The cases of a spin-orbit interaction due to random impurities and a localized Rashba interaction are considered. 
It is shown that the spin Hall effect has a broad response for the wave vector $q\lesssim 2\kf$ where $\kf $ is the Fermi wave vector. 
This fact confirms the local nature of the spin-charge conversion effects. 
}
\maketitle

\newcommand{\omegav}{{\bm \omega}}
\newcommand{\omegavE}{\omegav}
\newcommand{\omegaE}{\omega}
\newcommand{\sigmae}{\sigma_{\rm e}}
\newcommand{\Tv}{{\bm T}}
\newcommand{\Mvhat}{\hat{\Mv}}
\newcommand{\svhat}{\hat{\sv}}

\section{Introduction}

Spin Hall effect is widely used as a method to generate spin current by applying an electric field to nonmagnetic conductors \cite{Dyakonov71,Hirsch99}. 
It induces a spin accumulation at edges and interfaces and a torque in the case of a junction with a ferromagnet \cite{Manchon19}.
Conventionally, spin Hall effect has been theoretically argued in terms of spin Hall conductivity which is the ratio of the spin current and the applied electric field \cite{Hirsch99}. 
Although this description is convenient as an analogy with electric conductivity is applicable, it has a fundamental ambiguity of definition of spin current because of the nonconservation of spin current. 

The ambiguity can be avoided by discussing spin density, which is a physical observable, as in the original proposal of the spin Hall effect by Dyakonov and Perel \cite{Dyakonov71}.  
The spin Hall effect in the spin representation in the linear response regime is described by a response function of spin to an applied electric field, namely, a correlation function of spin and electric current \cite{TataraSH18}. 
A  mechanism to couple the spin and the electric current, such as the spin-orbit interaction and a magnetic structure, is essential to have a finite spin-electric current correlation function. 
Considering a spin-orbit interaction keeping the inversion symmetry, the spin-electric current correlation has vanishing uniform component, as it is the average of $s_i k_j$, a product of spin $\sv$ and wave vector $\kv$ ($i$ and $j$ represent directions), which is odd in inversion.
This is in contrast to the spin Hall conductivity, which is inversion-even and has a uniform component.
The result is natural physically as the spin accumulation generated by the spin Hall effect is not uniform, but is localized at the edges within the penetration length, determined by the rates of the  spin relaxation and diffusion \cite{Dyakonov71}. 
It was recently pointed out theoretically that  the leading contribution of the spin-electric current correlation is the term linear in the external wave vector $\qv$ (or spatial derivative), which corresponds to the inhomogeneous spin accumulation at edges \cite{TataraSH18}. 
When electron diffusion is neglected, the induced  spin density $\sv$ was shown to be proportional to the vorticity of the electric current or electric field $\Ev$ as 
\begin{align}
\sv= \lambda_{\rm sh}(\nabla\times\Ev), \label{shq0}
\end{align}
where $\lambda_{\rm sh}$ is a coefficient proportional to the strength of the spin orbit interaction and to the spin Hall angle of the conventional description. 
As was argued in Ref. \cite{Tatara21}, the spin Hall effect is a direct consequence of the spin-vorticity coupling \cite{Matsuo11}.

The expression (\ref{shq0}) clearly indicates that the spin Hall effect in terms of spin accumulation has vanishing bulk impact in the long wavelength limit, $\qv=0$. 
At large $\qv$ comparable to the Fermi wave vector $k_{\rm F}$, the correlation function is also small in the nonmagnetic case because the electron excitation is away from the Fermi energy in the spin-unpolarized case. 
The situation is different if electrons have different Fermi surfaces, like in ferromagnetic metal.
There, the Fermi surfaces for the electrons with spin up and down do not overlap and low energy excitation becomes possible at high $\qv$, in the broad regime of $|k_{{\rm F}+}-k_{{\rm F}-}|<q<k_{{\rm F}+}+k_{{\rm F}-}$ for the case of spherical Fermi surface with Fermi wave vectors $k_{{\rm  F}\pm}$ for spin $\pm$.
Such a low energy excitation with spin flip, known as the Stoner excitation, leads to a large weight for the magnetic susceptibility at large $q$.  
The low energy excitation is expected  to lead to an enhancement of the spin Hall response function in the Stoner regime in ferromagnets. 

Spin response function for the applied electric current was  discussed in the context of spin-orbit torque \cite{Manchon19} phenomenologically in Ref. \cite{Hals13} and density-functional calculation has been carried out  for  bilayers of a ferromagnet and a heavy metal \cite{Freimuth14,Freimuth15,Salemi21}.  
The torque is induced by the  spin accumulation at the interface generated by the spin Hall effect in the heavy metal, and thus the theories calculate a spin response function for the applied electric current. 
In the present paper, we explore analytically the finite-$q$ regime of the response function in a ferromagnetic metal with a quadratic free electron dispersion. 

\section{Linear response theory : Random impurity spin-orbit model}

\begin{figure}
 \includegraphics[width=0.4\hsize]{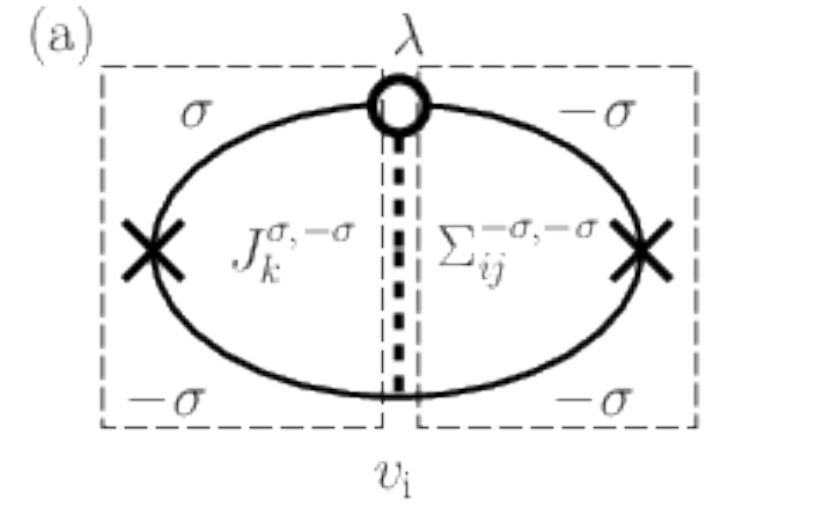}
 \includegraphics[width=0.4\hsize]{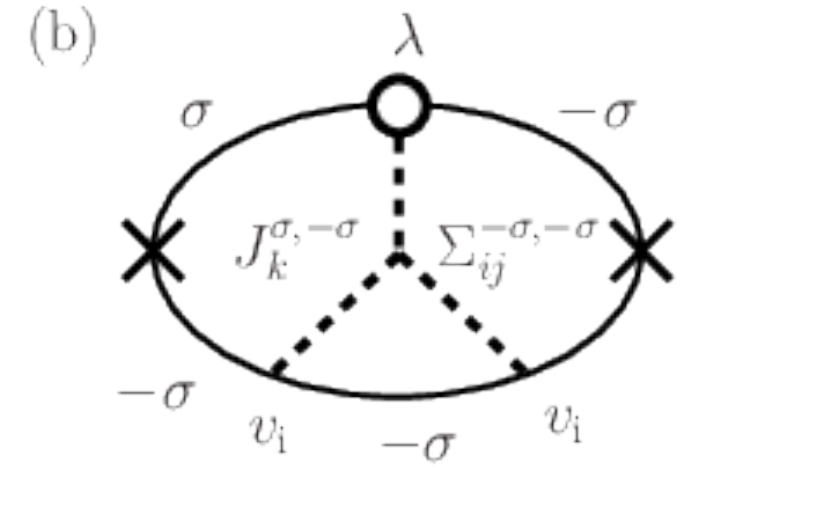}
 \caption{
 Feynman diagrams representing the response function of spin (left vertex) and electric current (right vertex). 
 Solid lines are electron Green's functions ($\sigma=\pm$ represents spin) and dotted lines denote impurity effects ($\vimp$ is the impurity potential)
 and the spin-orbit interaction is denoted by $\lambda$.  
 Diagram (b) is the dominant contributions at $q\sim0$ as has been known in the context of the anomalous Hall \cite{Crepieux01} and spin Hall effects \cite{TataraSH18}, while diagram (a) is dominant for large $q$.
\label{FIGdiag}}
\end{figure}

The spin Hall response function is a correlation function of spin and electric current, consisting of  
is  a spin vertex and a current vertex connected by the spin-orbit interaction, depicted in Fig. \ref{FIGdiag} \cite{TataraSH18}.
We consider ferromagnetic conductor with spin splitting $M$ with spin polarization along a unit vector $\Mvhat$.
The electron energy for wave vector $\kv$ and spin $\sigma=\pm$ measured from the Fermi energy $\ef$ is $\epsilon_{\kv\sigma}=\frac{k^2}{2m}-\sigma M-\ef$, $m$ being the mass. 
The spin-orbit interaction we consider is the one induced by impurities, represented by 
\begin{align}
 H_{\rm so} 
 &= \lambda \int d^3r  c^\dagger(\rv) [(\nabla v(\rv) \times \hat{\pv}) \cdot\sigmav] c(\rv)  \nnr
&= i\lambda \sum_{\kv\kv'} v_{\kv'-\kv} (\kv'\times\kv) \cdot c^\dagger_{\kv'} \sigmav c_{\kv} 
\end{align}
where $\lambda$ is a coupling constant, $v(\rv)=v_{\rm i}\sum_{\Rv_n}\delta(\rv-\Rv_n)$ is an impurity potential, where $v_{\rm i}$ and $\Rv_n$ are the strength of the impurity potential and the position of  $n$-th impurity, respectively, and the impurity scattering potential is $H_{\rm i}= \int d^3r v(\rv)  c^\dagger(\rv) c(\rv)$.
The impurity scattering induces an electron lifetime of elastic scattering, $\tau$, given by
$\tau^{-1}=2\pi\nu\nimp\vi^2$, where $\nimp$ is the impurity concentration and $\nu$ is the density of states of electron at the Fermi energy. 

We consider a spin density $\sv$ induced by an applied electric field $\Ev$ in the linear response regime.
The field is temporary static and spatially nonuniform as represented by a finite external wave vector $\qv$.
We first revisit the nonmagnetic  case argued in Ref. \cite{TataraSH18}, where the spin density response, 
$s^\alpha\equiv s_{i}^{{\rm sh},\alpha}E_i$ ($\alpha$ is spin direction and $i$ is spatial direction), was studied at $q\sim0$ for spin-unpolarized case. 
Near $q=0$, the spin response function was shown to be described by the process Fig. \ref{FIGdiag}(b), which reads 
\begin{align}
 s_{i}^{{\rm sh},\alpha}(\qv)&= 
 \frac{2i}{\pi } \frac{\lambda \nimp\vi^3}{m} \nu \epsilon_{\alpha jk} \Im
 [J_k(\qv) \Sigma_{ij}^{}(\qv) ]
 \label{skshdef}
 \end{align}
 where 
 \begin{align}
 J_i(\qv)
 &= 
 \frac{1}{V}\sum_{\kv}  
 k_i g_{\kv+\frac{\qv}{2}}^\ret
 g_{\kv-\frac{\qv}{2}}^\adv  
 \label{Jdef}
\end{align}
is the spin part (the factor $k_i$ arises from the spin-orbit interaction) and 
 \begin{align}
\Sigma_{ij}(\qv) \equiv
\frac{1}{V} \sum_{\kv}  
 k_i k_j  g_{\kv+\frac{\qv}{2}}^\ret g_{\kv-\frac{\qv}{2}}^\adv  
 \label{Sigmadef}
\end{align}
is a current-current response function ($V$ is a dimensionless volume). 
The retarded and advanced components of the electron free Green's functions at zero angular frequency are denoted by 
$g_{\kv}^\ret\equiv [-\epsilon_{\kv}+\frac{i}{2\tau}]^{-1}$ and $g_{\kv}^\adv$, respectively.
Near $q\sim 0$ \cite{TataraSH18}, 
we have $ J_i(\qv)=-i\frac{\pi}{3m}\nu\kf^2\tau^2 q_i+O(q^3)$ 
and 
$\Sigma_{ij}=  \delta_{ij} \frac{a^3 m^2}{e^2} \sigma_{{\rm e}} +O(q^2)$, where $\kf$ is the Fermi wave vector,  $\sigma_{{\rm e}} =e^2 n\tau/m$ is the Boltzmann conductivity, $n$ is the electron density, $a\sim \kf^{-1}$ is the lattice constant.
Thus we obtain  $\sv= \lambda_{\rm sh}(\nabla\times\Ev)$ with 
$\lambda_{\rm sh}=\frac{1}{9}\frac{\epsilon_{\rm so}}{\ef}(\ef\tau)^2 \frac{1}{\ef\kf^2}$ in the nonmagnetic case (assuming $\nu\sim\ef^{-1}$), where $\epsilon_{\rm so}\equiv \lambda\vi \kf^2$ and $\ef$ are the spin-orbit energy and the Fermi energy, respectively. 

Now we study the ferromagnetic case. 
We evaluate the electron spin density, 
\begin{align}
 s^\alpha(\rv,t) &= -i \tr [\sigma_\alpha {\cal G}^<(\rv,t,\rv,t)],
 \label{sdef}
\end{align}
where $\tr$ is the trace over spin, $\sigma_\alpha$ is the Pauli matrix with direction $\alpha$, ${\cal G}$ is the full electron Green's function and $^<$ denotes the lesser component. 
To explore spin-split case and general $\qv$, we go beyond the Fermi surface approximation employed in Ref. \cite{TataraSH18}, where the excitation represented by the product of the retarded and advanced Green's functions were considered.
In fact, the expression Eq. (\ref{Sigmadef}), which is commonly used for conductivity, diverges at large $k$ in the continuum model, but it gives a good approximation  concerning Fermi surface effects near $q=0$. 
The correct expression applicable to general cases is obtained straightforwardly by evaluating the lesser component in Eq. (\ref{sdef}). 
The contribution linear in a static applied electric field is written in the momentum representation as 
\begin{align}
 s^\alpha(\qv) &= i\frac{e}{V}\sum_{\kv}\sumom \sum_k k_k\lim_{\Omega\ra0} \tr [\sigma_\alpha G_{\kv+\frac{\qv}{2},\omega+\Omega}G_{\kv-\frac{\qv}{2},\omega}]^<
 A_k(\qv,\Omega)
\end{align}
where $\Av$ is a vector potential related to the electric field as $\Ev=-\dot{\Av}$ ($i\Omega  A_k(\qv,\Omega)=E_k(\qv)$) and 
the Green's function $G$ is without the external field. 
The linear response coefficient defined as $ s^\alpha =s_k^\alpha E_k $  is therefore 
\begin{align}
 s_k^\alpha(\qv) &= i\frac{e}{V}\sum_{\kv}\sumom \sum_k k_k\lim_{\Omega\ra0}\frac{1}{\Omega} \tr [\sigma_\alpha G_{\kv+\frac{\qv}{2},\omega+\Omega}G_{\kv-\frac{\qv}{2},\omega}]^<
\end{align}
The static limit in the case of static interactions is given by the contribution of 
\begin{align}
[G_{\kv+\frac{\qv}{2},\omega+\Omega}G_{\kv-\frac{\qv}{2},\omega}]^<
=\Omega f'(\omega) \lt[G_{\kv+\frac{\qv}{2},\omega}^\ret G_{\kv-\frac{\qv}{2},\omega}^\adv -\frac{1}{2}(G_{\kv+\frac{\qv}{2},\omega}^\ret G_{\kv-\frac{\qv}{2},\omega}^\ret +G_{\kv+\frac{\qv}{2},\omega}^\adv G_{\kv-\frac{\qv}{2},\omega}^\adv ) \rt]+O(\Omega^2)
\end{align}
where $f(\omega)\equiv (e^{\beta\omega}+1)^{-1}$ is the Fermi distribution function ($\beta\equiv (\kb T)^{-1}$ is the inverse temperature, $\kb$ being the Boltzmann constant). 
The Green's function $G$ is calculated perturbatively taking account of the spin-orbit interaction. 
The impurity scattering is included to compensate the momentum change due to the spin-orbit interaction. 
In the case of uniform ($q=0$) response function, the diagram of Fig. \ref{FIGdiag}(b) including the impurity to the third order is dominant \cite{Crepieux01,TataraSH18}, while the second-order diagram is also important at finite $q$. 
The second-order contribution is 
\begin{align}
 s_{k}^{(2) \alpha}(\qv)
 &=-i\frac{e}{\pi}\frac{\lambda \nimp \vi^2}{m}\frac{1}{V^2}  \sum_{\kv\kv'}  k'_k 
\lt[\lt(\kv'+\frac{\qv}{2}\rt)\times\lt(\kv+\frac{\qv}{2}\rt)\rt]_\beta \nnr
& \times 
\Re \tr\biggl[
\sigma_\alpha g_{\kv+\frac{\qv}{2}}^\ret \sigma_\beta g_{\kv'+\frac{\qv}{2}}^\ret
 g_{\kv'-\frac{\qv}{2}}^\adv  g_{\kv-\frac{\qv}{2}}^\adv 
 -\sigma_\alpha g_{\kv+\frac{\qv}{2}}^\ret \sigma_\beta g_{\kv'+\frac{\qv}{2}}^\ret
 g_{\kv'-\frac{\qv}{2}}^\ret  g_{\kv-\frac{\qv}{2}}^\ret \biggr]
 \label{s2def}
\end{align}
where $g_{\kv\sigma}^\ret\equiv [-\epsilon_{\kv\sigma}+\frac{i}{2\tau}]^{-1}$ and $g_{\kv\sigma}^\adv$ are the Free Green's functions at zero-frequency.

The trace for spin is calculated using
\begin{align}
\tr[\sigma^\alpha A \sigma^\beta B]  
&=
(\delta_{\alpha\beta} -\delta_{\alpha z} \delta_{\beta z})
\sum_{\sigma} A_\sigma B_{-\sigma}
+\delta_{\alpha z} \delta_{\beta z}
\sum_{\sigma} A_\sigma B_{\sigma} 
- i\epsilon_{\alpha\beta z} \sum_{\sigma} \sigma A_\sigma B_{-\sigma} 
\label{traceformula}
\end{align}
for $2\times2$ diagonal matrices, 
$A\equiv\lt(\begin{array}{cc} A_+ & 0 \\ 0 & A_- \end{array} \rt)$ and $B\equiv\lt(\begin{array}{cc} B_+ & 0 \\ 0 & B_- \end{array} \rt)$. 
As was argued in Ref. \cite{Tatara21}, we have three components of spin accumulation in ferromagnets, 
$ s_{k}^{\alpha}=  s_{k}^{{\rm sh},\alpha}+ s_{k}^{{\perp},\alpha}+ s_{k}^{{\rm ad},\alpha}$, 
where $s_{k}^{{\rm sh},\alpha}$ is the spin Hall contribution proportional to  $\qv\times\Ev$ (vorticity in the $\qv$-representation), $s_{k}^{{\perp},\alpha}$ is the perpendicular component proportional to $\Mv\times(\qv\times\Ev)$ and $s_{k}^{{\rm ad},\alpha}$ is the adiabatic component parallel to $\Mv$.  
(Note that $\qv\times\Ev$ at finite $q$ regime with extra dependence on $\qv$ does not exactly corresponds to the vorticity in the real space, $\nabla\times \Ev$.)
We focus on the nonadiabatic components $s_{k}^{{\rm sh},\alpha}$ and $s_{k}^{{\perp},\alpha}$.  
The contribution second order of impurity, Eq. (\ref{s2def}), is 
\begin{align}
 s_{k}^{{\rm sh}(2) \alpha}(\qv)
 &= -i\frac{e}{\pi}\frac{\lambda\nimp \vi^2}{m} \epsilon_{\alpha ij} \sum_{\sigma}
 \Re \lt[J_j^{{\rm (ra)},\sigma,-\sigma}(\qv) \Sigma_{ik}^{{\rm (ra)},-\sigma,-\sigma}-J_j^{{\rm (rr)},\sigma,-\sigma}(\qv) \Sigma_{ik}^{{\rm (rr)},-\sigma,-\sigma}\rt]
\nnr
 s_{k}^{{\perp}(2) \alpha}(\qv)
 &=-i\frac{e}{\pi}\frac{\lambda \nimp\vi^2}{m} 
 \epsilon_{\alpha\beta z}\epsilon_{\beta ij} \sum_{\sigma} \sigma 
\Im  \lt[J_j^{{\rm (ra)},\sigma,-\sigma}(\qv) \Sigma_{ik}^{{\rm (ra)},-\sigma,-\sigma}-J_j^{{\rm (rr)},\sigma,-\sigma}(\qv) \Sigma_{ik}^{{\rm (rr)},-\sigma,-\sigma}\rt]
\label{spins}
\end{align}
Here (A, B=r, a)
\begin{align}
 J_j^{{\rm (AB)},\sigma,-\sigma}(\qv) 
  &\equiv \frac{1}{V} \sum_{\kv}  \lt(\kv+\frac{\qv}{2}\rt)_j  g_{\kv+\frac{\qv}{2},\sigma}^{\rm A} g_{\kv-\frac{\qv}{2},-\sigma}^{\rm B} \nnr
 \Sigma_{ik}^{{\rm (AB)},\sigma,\sigma} 
 &\equiv \frac{1}{V} \sum_{\kv'}  k'_k 
\lt(\kv'+\frac{\qv}{2}\rt)_i  g_{\kv'+\frac{\qv}{2},\sigma}^{\rm A}
 g_{\kv'-\frac{\qv}{2},\sigma}^{\rm B},
\end{align}
are extensions of Eqs. (\ref{Jdef}) (\ref{Sigmadef}) to spin-polarized case with finite $q$.
As is easily seen from symmetry, 
\begin{align}
 J_j^{{\rm (AB)},\sigma,-\sigma}
 &=\hat{q}_j \overline{J}^{{\rm (AB)},\sigma,-\sigma}\nnr
 \Sigma_{ik}^{{\rm (AB)},\sigma,\sigma} 
 &= \delta_{ik} \overline{\Sigma}^{{\rm (AB)},\sigma,\sigma}+\hat{q}_i \hat{q}_k \overline{\Sigma}^{{\rm (AB)}(2),\sigma,\sigma}
 \label{qdep}
\end{align}
where 
\begin{align}
\overline{J}^{{\rm ra},\sigma,-\sigma}
 & \equiv 
 \frac{1}{V} \sum_{\kv}  \lt(k_z+\frac{q}{2}\rt)  g_{\kv+\frac{\qv}{2},\sigma}^\ret g_{\kv-\frac{\qv}{2},-\sigma}^\adv 
 \nnr
 \overline{\Sigma}^{{\rm ra},\sigma,\sigma} 
 & \equiv  \frac{1}{V} \sum_{\kv'}  \frac{(k')^2}{3}
 g_{\kv'+\frac{\qv}{2},\sigma}^\ret g_{\kv'-\frac{\qv}{2},\sigma}^\adv ,
 & 
 \overline{\Sigma}^{{\rm ra}(2),\sigma,\sigma} 
 & \equiv  \frac{1}{V} \sum_{\kv'}  
 \lt[ k'_z \lt(k'_z+\frac{q}{2}\rt) -\frac{(k')^2}{3}\rt]
 g_{\kv'+\frac{\qv}{2},\sigma}^\ret g_{\kv'-\frac{\qv}{2},\sigma}^\adv ,
 \label{JSigmadefs}
\end{align}
and similarly for r and a components, 
where the $z$ axis of $\kv$ here are chosen along $\hat{\qv}$.   
Eq. (\ref{spins}) therefore read
\begin{align}
 s_{k}^{{\rm sh}(2) \alpha}(\qv)
 &= -i
 \overline{\lambda}_{\rm sh}\epsilon_{\alpha kj} \hat{q}_j 
  \overline{s}^{{\rm sh}(2)} (q)\nnr
 s_{k}^{{\perp}(2) \alpha}(\qv)
 &=-i  \overline{\lambda}_{\rm sh}
 \epsilon_{\alpha\beta z}\epsilon_{\beta kj}  \hat{q}_j  \overline{s}^{{\perp}(2)} (q)
\label{spins2}
\end{align}
where 
\begin{align}
 \overline{\lambda}_{\rm sh} 
 \equiv & \frac{e}{\pi}\frac{\lambda\nimp \vi^2}{m} \frac{\kf^3}{\ef^4} 
 = \frac{e}{\pi}\frac{\epsilon_{\rm so}}{\ef} \frac{\nimp\vimp}{\ef}\frac{1}{\kf \ef} 
\end{align}
and 
\begin{align}
\overline{s}^{{\rm sh}(2)} (q)
\equiv & 
\frac{\ef^4}{\kf^3} 
\sum_{\sigma}  \Re \lt[\overline{J}^{{\rm (ra)}\sigma,-\sigma} \overline{\Sigma}^{{\rm (ra)}-\sigma,-\sigma} 
 - \overline{J}^{{\rm (rr)}\sigma,-\sigma} \overline{\Sigma}^{{\rm (rr)}-\sigma,-\sigma} \rt]
\nnr
\overline{s}^{{\perp}(2)} (q)
\equiv &
\frac{\ef^4}{\kf^3} 
 \sum_{\sigma} \sigma\Im \lt[\overline{J}^{{\rm (ra)}\sigma,-\sigma} \overline{\Sigma}^{{\rm (ra)}-\sigma,-\sigma}
 -\overline{J}^{{\rm (rr)}\sigma,-\sigma} \overline{\Sigma}^{{\rm (rr)}-\sigma,-\sigma}\rt]
\label{spincoefficients2}
\end{align}
are dimensionless coefficients. 
Equation (\ref{spins2}) indicates that directions of spin polarizations are  
$ \sv^{{\rm sh}(2)}(\qv) \propto (\qv\times \Ev)$ and $ \sv^{{\perp}(2)}(\qv) \propto \Mvhat\times (\qv\times \Ev)$.

\begin{figure}
\centering
 \includegraphics[width=0.45\hsize]{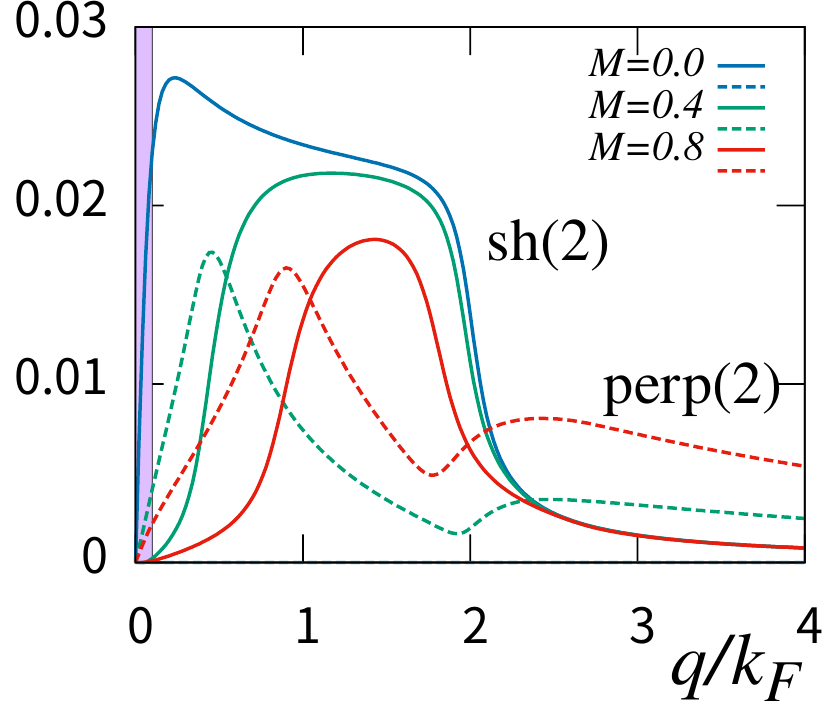}
 \includegraphics[width=0.45\hsize]{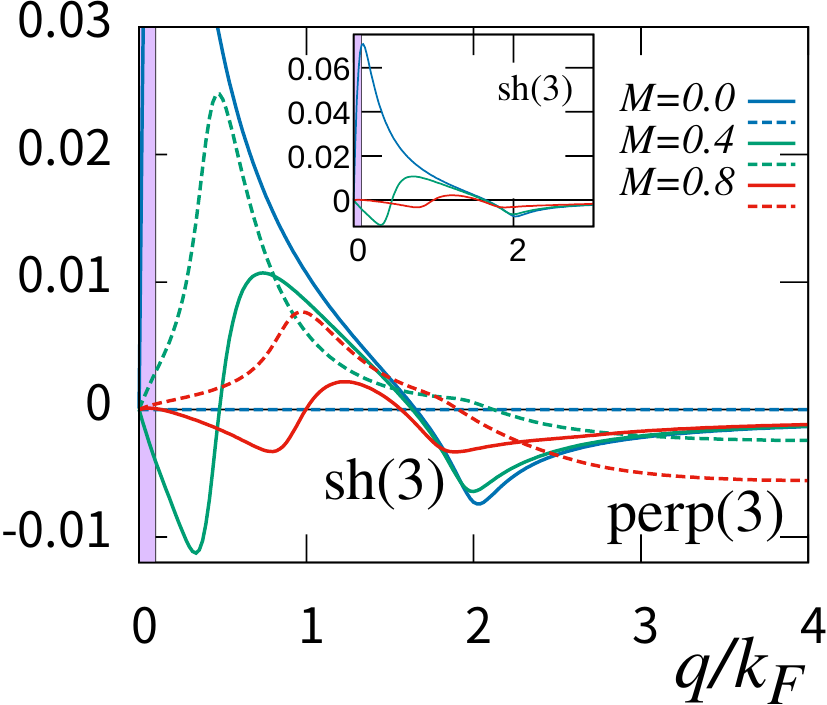}
 \caption{The amplitude of spin Hall response functions 
 $ \overline{s}^{{\rm sh}(n)}$ and $ \overline{s}^{{\perp}(n)}$ for the second and third-order of impurity ($n=2,3$ respectively) plotted 
 as functions of wave vector $q$ normalized by $\kf$. Spin splitting is $M=0, 0.4, 0.8$ in unit of $\ef$ and the damping constant is $\eta/\ef=0.1$.
Shaded regime is the contribution for the uniform spin Hall effect for the case of spin diffusion length of $\ell_s=10/\kf$. 
\label{FIGplotS2}}
\end{figure}

A key factor determining the spin Hall response is the spin part function 
$\overline{J}^{{\rm AB},\sigma,-\sigma}$, that involves the two opposite spins. 
This function has a large contribution from $\kv$ satisfying $\epsilon_{\kv,\sigma}=0$ (redefining $\kv+\frac{\qv}{2}$ as $\kv$ in Eq. (\ref{JSigmadefs})) and 
$\epsilon_{\kv+\qv,-\sigma}=0$, i.e., at $k=k_{{\rm F}\sigma}$ and $|\kv+\qv|=k_{{\rm F},-\sigma}$.
These conditions are satisfied when $|k_{{\rm F}+}-k_{{\rm F}-}|\leq q \leq k_{{\rm F}+}+k_{{\rm F}-}$, namely, in the Stoner regime for the external wave vector $q$. 
In the regime, the amplitude of the function is the order of $\kf/\ef^2$ regardless of the value of $\eta\equiv(2\tau)^{-1}$. 
Considering $q=O(\kf)$ in the Stoner regime, 
the spin density has therefore an enhancement factor of the order of $\kf \ell_s$ compared to the conventional spin Hall effect near $q=0$ (Eq. (\ref{shq0})), where $\ell_s$ is the length of spin propagation (spin diffusion length), related to $q$ as $q(=-i\nabla)\sim 1/\ell_s $ (Fig. \ref{FIGplotS2}).
In contrast, function $ \overline{\Sigma}^{{\rm AB},\sigma,\sigma} $ is not sensitive to spin-splitting. 
The real part of $ \overline{\Sigma}^{{\rm ra},\sigma,\sigma} $ has a sharp peak at $q=0$  of magnitude and width proportional to $\tau$ and $\tau^{-1}$, respectively, while the imaginary part is small.  

The contribution third order of the impurity, Fig. \ref{FIGdiag}(b), is similarly obtained as 
\begin{align}
 s_{k}^{{\rm sh}(3) \alpha}(\qv)
 &= -i \overline{\lambda}_{\rm sh} \pi\frac{\vi}{\ef} \epsilon_{\alpha ij}\hat{q}_j
\overline{s}^{{\rm sh}(3)} (q)
\nnr
 s_{k}^{{\perp}(3) \alpha}(\qv)
 &=
 -i \overline{\lambda}_{\rm sh}  \pi\frac{\vi}{\ef}\epsilon_{\alpha\beta z}\epsilon_{\beta ij} \hat{q}_j
\overline{s}^{{\perp}(3)} (q)
\end{align}
where
\begin{align}
\overline{s}^{{\rm sh}(3)} (q)
 \equiv & 
-\frac{\ef^5}{\kf^3} 
 \sum_\sigma \nu_{-\sigma}\Im  \lt[ \overline{J}^{{\rm (ra)}\sigma,-\sigma} \overline{\Sigma}^{{\rm (ra)}-\sigma,-\sigma}
 + \overline{J}^{{\rm (rr)}\sigma,-\sigma} \overline{\Sigma}^{{\rm (rr)}-\sigma,-\sigma} \rt]
\nnr %
\overline{s}^{{\perp}(3)} (q)
 \equiv & 
 \frac{\ef^5}{\kf^3}
 \sum_\sigma \sigma  \nu_{-\sigma} \Re 
\lt[\overline{J}^{{\rm (ra)}\sigma,-\sigma} \overline{\Sigma}^{{\rm (ra)}-\sigma,-\sigma}
  +  \overline{J}^{{\rm (rr)}\sigma,-\sigma} \overline{\Sigma}^{{\rm (rr)}-\sigma,-\sigma} \rt]\label{spins32}
\end{align}
Compared to the second-order process, the third-order one includes one more self-energy, which is dominantly pure imaginary ($\pm i\pi\nu$), and thus the roles of the real and imaginary parts of the response function are interchanged \cite{Crepieux01}. 
At $q=0$, the current-current correlation function 
$ \Sigma_{ik}^{{\rm (ra)},\sigma,\sigma} $ has large real part with vanishing imaginary part, while the current function 
$ J_j^{{\rm (ra)},\sigma,-\sigma}$ is pure imaginary proportional to $q_j$. We have therefore large imaginary part of the product of the two functions, resulting in a large third-order contribution, 
$\overline{s}^{{\rm sh}(3)}$, at $q\sim0$ (Fig. \ref{FIGplotS2}). (The same mechanism works for the anomalous Hall effect \cite{Crepieux01}.) 
The feature is different  at finite $q$ in the ferromagnetic case,  and the second-order diagram becomes dominant. 

The plot of the amplitudes $ \overline{s}^{{\rm sh}(n)}(\qv)$ and $ \overline{s}^{{\perp}(n)}(\qv)$ ($n=2,3$) 
are shown in Fig. \ref{FIGplotS2}.
The perpendicular components $\overline{s}^{{\perp}(n)}(\qv)$ are small compared to  $ \overline{s}^{{\rm sh}(n)}(\qv)$. 
It is seen that the second-order spin Hall contribution  $ \overline{s}^{{\rm sh}(2)}$ is dominant for a broad regime of $q\lesssim 2\kf$ with a height insensitive to $\eta$. 
The spin Hall effect has thus a broad response up to a high-wave vector regime in ferromagnets. 
This feature indicates that  the induced spin accumulation has a localized nature, namely, a local nature of the spin-charge conversion. 
Conventional spin Hall effect in nonmagnets corresponds to $ \overline{s}^{{\rm sh}(3)}$ near $q=0$ with a high peak $\propto \tau$.

\subsection{Discussion}
In Ref. \cite{Salemi21}, the spin magnetoelectric susceptibility $\chi_{s}$ in a junction was defined as $s=\chi_s E$ and its $q=0$ contribution was theoretically discussed. 
Their spin density proportional to $\Ev\times \uv$  (called $\Ev_\perp$ component), where $\uv$ is the normal vector to the interface, corresponds to the $q\ra0$ limit of our $\sv^{{\rm sh}}$, and  density proportional to $\Mv\times(\Ev\times \uv)$ (called $\Mv_\perp$) corresponds to our $\sv^{{\perp}}$. Thus the susceptibilities in our notation read 
\begin{align}
\chi_{s}^{\Ev_\perp} &= 
\overline{\lambda}_{\rm sh} \lim_{q\ra0} \frac{1}{q}
\lt[  \overline{s}^{{\rm sh}(2)} (q)+  \frac{\pi\vi}{\ef} \overline{s}^{{\rm sh}(3)} (q) \rt] \nnr
\chi_{s}^{\Mv_\perp} &= 
\overline{\lambda}_{\rm sh} \lim_{q\ra0} \frac{1}{q}
\lt[  \overline{s}^{{\perp}(2)} (q)+  \frac{\pi\vi}{\ef} \overline{s}^{{\perp}(3)} (q) \rt] \nnr
\end{align}
The magnitude of $\frac{e}{\pi}\frac{1}{\kf \ef} $  in $\overline{\lambda}_{\rm sh}$ is evaluated for $\kf^{-1}=1\AA$ and $\ef=1$ eV as $\frac{e}{\pi}\frac{1}{\kf \ef} =3\times 10^{-11}$ m/V. 
Considering the numerical factor of $\overline{s}^{{\rm sh}(2)} \sim 0.02$ in the Stoner regime (Fig. \ref{FIGplotS2}), our estimate of the susceptibility in the Stoner regime is $\chi_s \sim 6\times 10^{-13} \times \frac{\epsilon_{\rm so}}{\ef} \frac{\nimp\vimp}{\ef}$ m/V.
At $q=0$, the susceptibility is enhanced by a factor of $\eta^{-1}$ due to a peak of $\Re\overline{\Sigma}^{{\rm ra}}$. 
In the case of Pt heavy metal, $\chi_s\sim 10^{-11}$ m/V at $q=0$ is predicted in Ref. \cite{Salemi21}. 
Our estimate based on a simplified model therefore agrees by the order of magnitudes with Ref. \cite{Salemi21} if
$ \frac{\epsilon_{\rm so}}{\ef} \frac{\nimp\vimp}{\ef}\sim1$ is assumed for heavy metals. 
Considering the fact that spin-orbit torque is highly efficient for magnetization switching \cite{Manchon19},  spin Hall effect in the Stoner regime is expected to induce various finite-$q$ magnetization dynamics.

\section{Rashba spin-orbit interaction}
Spin-orbit interaction due to random impurities conserves momentum, and thus the wave vector $\qv$ driving the finite-$\qv$ spin Hall effect needs to be given by the applied electric field. 
Due to the momentum conserving response function, the  spin accumulation and the  electric field related by the direct and inverse spin Hall effects have the same wave length.   
It is not, however, easy to access by electrical methods the Stoner regime, which corresponds to a high wave vector regime of the order of lattice constant in ferromagnetic metals.
Doped ferromagnetic semiconductors \cite{Konig00}  having an orders of magnitude longer wave length is one promising system for finite-$q$ spin Hall effect. 
From the viewpoint of finite $q$, use of surface and interface is of special interest, as finite wave vector transfer is allowed due to the inversion symmetry breaking.

\begin{figure}
\centering
 \includegraphics[width=0.3\hsize]{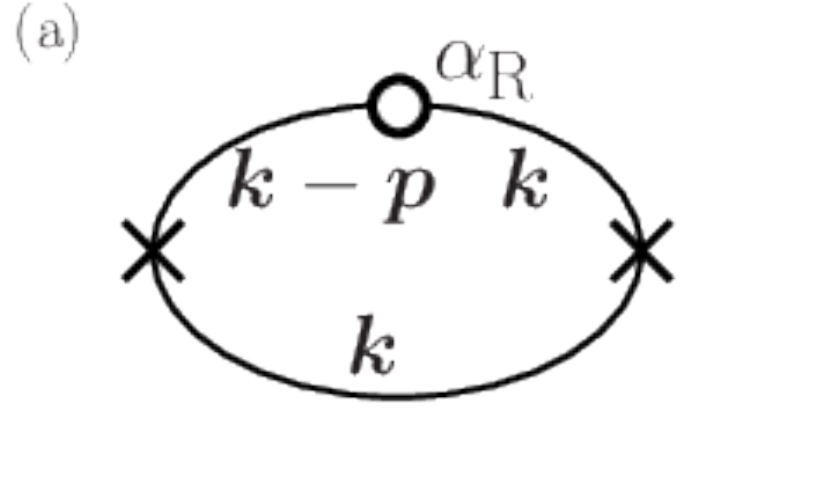}
 \includegraphics[width=0.3\hsize]{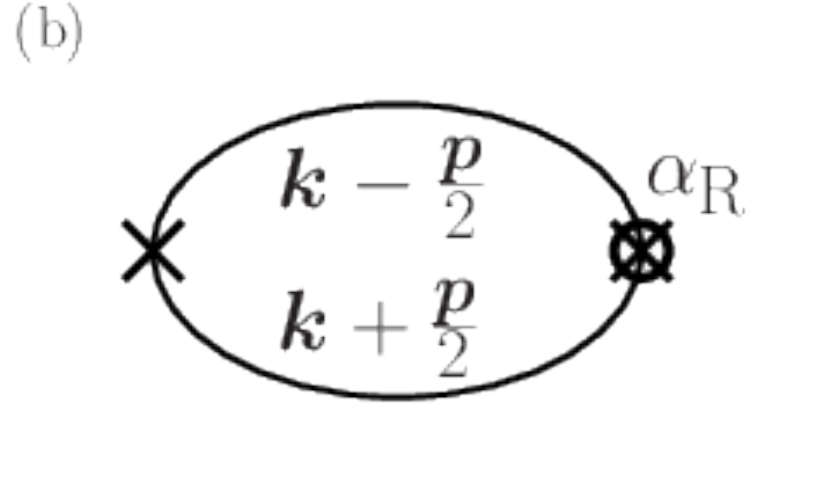}
 \caption{ Diagrams contributing to the response function of electron spin and current at the linear order of the Rashba spin-orbit interaction denoted by circles. Diagram (b) is the one with a current vertex modified by the Rashba interaction. 
 \label{FIGdiagRashba}}
\end{figure}

We consider  the Rashba spin-orbit interaction, whose Hamiltonian is 
\begin{align}
 H_{\rm R} &= \int d\rv \alpha_{{\rm R},i}\epsilon_{ijk} c^\dagger \hat{p}_j \sigma_k c
\end{align}
where $\alphav_{\rm R}$ is the Rashba vector. 
We consider the case where $\alphav_{\rm R}$ is along the $z$ axis and localized at the surface chosen as $z=0$;
Namely, $\alphav_{\rm R}(\rv)=\alpha_{\rm R}(z)\zvhat $ with a coefficient $\alpha_{\rm R}(z)=\overline{\alpha}_{\rm R} a\delta(z)$ ($a$ is the lattice constant introduced for using conventional unit). 
In the momentum representation, we have
\begin{align}
 H_{\rm R} &= -\sum_{\pv\kv} \alpha_{\rm R}(\pv)\epsilon_{ijk} k_j c^\dagger_{\kv+\frac{\pv}{2}} \sigma_k c_{\kv-\frac{\qv}{2}}
\end{align}
The interaction modifies the electron current density as 
\begin{align}
 j_\alpha(\rv) &= e^{-i(\qv-\pv)\cdot\rv} c^\dagger_{\kv+\qv}\lt[\frac{k_\alpha}{m}\delta_{\pv,0}-\epsilon_{\alpha lm} \alpha_{{\rm R},l}(\pv)\sigma_m \rt]  c_{\kv}
\end{align}
We consider the Rashba interaction to the linear order to evaluate the response function of electron spin and current for a uniform applied electric field.
The Feynman diagrams representing the response function are in Fig. \ref{FIGdiagRashba}.

\begin{figure}
\centering
 \includegraphics[width=0.5\hsize]{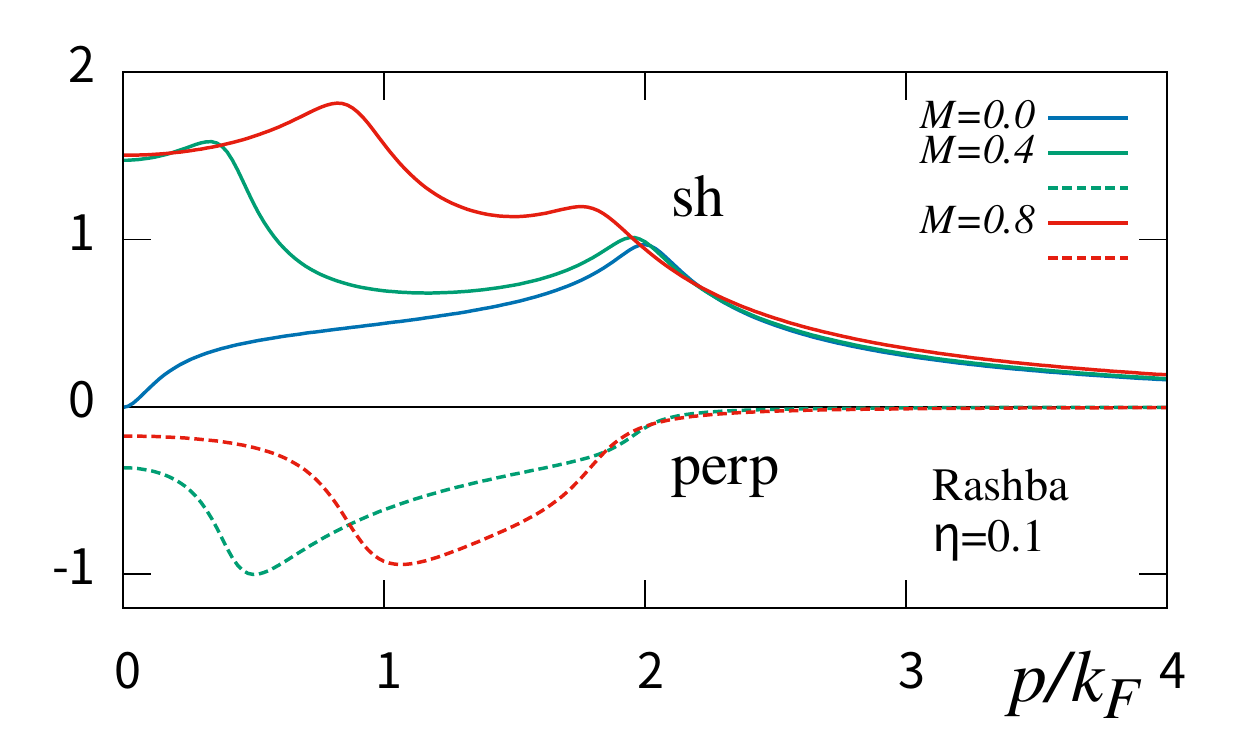}
 \caption{ The response functions  $ s_{{\rm R},i}^{{\rm sh},\alpha} $ and $ s_{{\rm R},i}^{\perp,\alpha}$ of electron spin and current at the linear order of the Rashba spin-orbit interaction as function of momentum transfer $p$ normalized by $\kf$. 
 The response for $M=0$ near $p=0$  is qualitatively different from those for $M\neq0$. Damping constant is $\eta/\ef=0.1$.
 \label{FIGsRashba}}
\end{figure}

The response functions defined by the ratio of induced spin in the direction $\alpha$ to the applied electric field in the $i$ direction is
((a) and (b) correspond to Fig. \ref{FIGdiagRashba}.)
\begin{align}
 s_{{\rm R},i}^{{\rm (a)},\alpha}(p) &=-2 \epsilon_{jl\beta} \sum_{p} \alpha_{{\rm R},j}(p) \frac{1}{m}\frac{1}{V}\sum_{\kv} \lt(k-\frac{p}{2}\rt)_l k_i 
 \Re \tr \lt[ \sigma_\alpha g^\ret_{\kv-\pv} \sigma_\beta g^\ret_{\kv} g^\adv_{\kv}\rt] \nnr
  s_{{\rm R},i}^{{\rm (b)},\alpha}(p) &= \epsilon_{ij\beta} \sum_{p} \alpha_{{\rm R},j}(p) \frac{1}{V}\sum_{\kv} 
 \Re \tr \lt[ \sigma_\alpha g^\ret_{\kv-\frac{\pv}{2}} \sigma_\beta  g^\adv_{\kv+\frac{\pv}{2}}\rt] 
\end{align}
Evaluating the trace for the spin, we obtain the spin Hall and anomalous Hall components, 
 $s_{\rm R}^{{\rm (n)},\alpha}=s_{\rm R}^{{\rm sh}{\rm (n)},\alpha}+s_{\rm R}^{{\perp}{\rm (n)},\alpha}$ (n=a,b). 
In the geometry we consider, $ \alpha_{{\rm R},j}\propto \delta_{jz}$,  and 
we obtain the sum of contributions (a) and (b) as 
\begin{align}
 s_{{\rm R},i}^{{\rm sh},\alpha}(p) 
  &= 2 \epsilon_{ij\alpha} \sum_{p} \alpha_{{\rm R},j}(p) \sum_{\sigma}
 \Re \biggl[J_{\rm R}^\sigma(p) + \Sigma_{\rm R}^\sigma(p) \biggr]\nnr 
 s_{{\rm R},i}^{\perp,\alpha}(p)
  &= 2 \epsilon_{ij\beta} \epsilon_{\alpha\beta\gamma}\hat{M}_\gamma \sum_{p} \alpha_{{\rm R},j}(p)  \sum_{\sigma} \sigma \Im \biggl[J_{\rm R}^\sigma(p) + \Sigma_{\rm R}^\sigma(p)   \biggr] 
\end{align}
where 
\begin{align}
J_{\rm R}^\sigma(p) &\equiv 
\frac{1}{V}\sum_{\kv}  \frac{k^2}{3m} g^\ret_{\kv-\pv,\sigma} g^\ret_{\kv,-\sigma} g^\adv_{\kv,-\sigma} \nnr
\Sigma_{\rm R}^\sigma(p) &\equiv 
       \frac{1}{V}\sum_{\kv} g^\ret_{\kv-\frac{\pv}{2},\sigma} g^\adv_{\kv+\frac{\pv}{2},-\sigma} 
\end{align}
As was shown in Ref. \cite{TataraSH18}, the spin Hall contribution $s_{{\rm R},i}^{{\rm sh}}(p)$ vanishes linearly at $p=0$ if $M=0$. 
Numerically calculated response function is plotted in Fig. \ref{FIGsRashba}.
The Rashba system has a broad response for $q\lesssim 2\kf$, meaning that the real-space distribution of the induced spin has a localized nature.  
This fact is consistent with the local spin-charge conversion picture of the Rashba-Edelstein effect.

The broad spin response of the localized Rashba interaction would be useful for the emission of finite-$q$ magnons. 

\section{Discussion}
The electron spin density induced by the spin Hall effect couples to the magnetization via the exchange coupling (with coupling constant $J_{sM}$).
The coupling leads to a so called the spin-orbit torque \cite{Manchon19} and drives magnetization and induces magnon emission.
Writing the transverse induced spin as $\sv_{\rm ind}$, i.e., 
$\sv_{\rm ind}=\sv^{\rm sh}+\sv^\perp$ and $\sv_{\rm ind}=\sv^{\rm sh}_{\rm R}+\sv^\perp_{\rm R}$ for spin-orbit interaction due to random impurities and the Rashba interaction, respectively, the exchange coupling Hamiltonian is $H_{sM}\equiv -J_{sM}\sv_{\rm ind}\cdot\Mv$. 
In terms of the Holstein-Primakoff boson operators, $a$ and $a^\dagger$, the magnetization (with quantum axis along $\hat{\zv}$) is
$\Mv=M\hat{\zv}+ \sqrt{\frac{M}{2}}(a^\dagger+a,i(a^\dagger-a),0)+O((a,a^\dagger)^2)$ to the linear order. 
The induced transverse spin  thus  couples linearly to the magnon operators as 
\begin{align}
 H_{sM}= -J_{sM}\sqrt{\frac{M}{2}}(s^{+}_{\rm ind}a^\dagger+s^{-}_{\rm ind}a) 
\end{align}
where $s^{\pm}_{\rm ind} \equiv s^{x}_{\rm ind}\pm is^{y}_{\rm ind}$. 
The wave vector of the induced spin is therefore transferred directly to magnons, and this feature would be useful for magnon manipulations \cite{Chumak15}.  
Experimentally, magnon energy can be chosen by use of an alternating current, and the magnon emission amplitude in this case is expected to be proportional to  $s^{\pm}_{\rm ind} (q_{\rm m})$, where $q_{\rm m}$ is the wave vector for the excited magnon.


\section{Summary}
We have studied the spin Hall effect in ferromagnets at finite wave vector by calculating the linear response function of spin and electric current for the spin-orbit interactions due to random impurity and the Rashba interaction.
It was shown that the response function shows a broad response in the regime $q\lesssim 2\kf$ due to the gapless Stoner excitation,  
suggesting a local nature of the spin-charge conversion effects. 
For experimental study of finite-$q$ regime, dilute magnetic semiconductors or use of STM would be suitable.  

\acknowledgements
The author thank E. Saitoh for valuable discussion. 
This study was supported by
a Grant-in-Aid for Scientific Research (B) (No. 17H02929 and No. 21H01034) from the Japan Society for the Promotion of Science.


\bibliographystyle{jpsj}

\begin{thebibliography}{10}

\bibitem{Dyakonov71}
M.~Dyakonov and V.~I. Perel: Sov. Phys. JETP Lett. {\bfseries 13} (1971) 467.

\bibitem{Hirsch99}
J.~E. Hirsch: Phys. Rev. Lett. {\bfseries 83} (1999) 1834.

\bibitem{Manchon19}
A.~Manchon, J.~\ifmmode~\check{Z}\else \v{Z}\fi{}elezn\'y, I.~M. Miron,
  T.~Jungwirth, J.~Sinova, A.~Thiaville, K.~Garello, and P.~Gambardella: Rev.
  Mod. Phys. {\bfseries 91} (2019) 035004.

\bibitem{TataraSH18}
G.~Tatara: Phys. Rev. B {\bfseries 98} (2018) 174422.

\bibitem{Tatara21}
G.~Tatara: Phys. Rev. B {\bfseries 104} (2021) 184414.

\bibitem{Matsuo11}
M.~Matsuo, J.~Ieda, E.~Saitoh, and S.~Maekawa: Phys. Rev. B {\bfseries 84}
  (2011) 104410.

\bibitem{Hals13}
K.~M.~D. Hals and A.~Brataas: Phys. Rev. B {\bfseries 88} (2013) 085423.

\bibitem{Freimuth14}
F.~Freimuth, S.~Bl\"ugel, and Y.~Mokrousov: Journal of Physics: Condensed
  Matter {\bfseries 26} (2014) 104202.

\bibitem{Freimuth15}
F.~Freimuth, S.~Bl\"ugel, and Y.~Mokrousov: Phys. Rev. B {\bfseries 92} (2015)
  064415.

\bibitem{Salemi21}
L.~Salemi, M.~Berritta, and P.~M. Oppeneer: Phys. Rev. Materials {\bfseries 5}
  (2021) 074407.

\bibitem{Crepieux01}
A.~Cr\'epieux and P.~Bruno: Phys. Rev. B {\bfseries 64} (2001) 014416.

\bibitem{Konig00}
J.~K\"onig, H.-H. Lin, and A.~H. MacDonald: Phys. Rev. Lett. {\bfseries 84}
  (2000) 5628.

\bibitem{Chumak15}
A.~V. Chumak, V.~I. Vasyuchka, A.~A. Serga, and B.~Hillebrands: Nat Phys
  {\bfseries 11} (2015) 453.
\newblock Review.

\end{thebibliography}

\end{document}